\documentclass[12pt]{article}
\usepackage{amssymb}
\usepackage{amsfonts}
\usepackage{color}
\usepackage{amsmath}
\usepackage[margin = 1in]{geometry}
\usepackage{natbib}
\usepackage[doublespacing]{setspace}
\usepackage{amsthm}
\theoremstyle{plain}
\usepackage[flushleft]{threeparttable}
\usepackage{adjustbox}
\usepackage{caption}
\usepackage{subcaption}
\usepackage{hyperref}
\usepackage{arydshln}
\usepackage{float}
\usepackage{adjustbox}
\usepackage[flushleft]{threeparttable}
\usepackage{lipsum}
\usepackage{pdflscape}
\usepackage{graphicx}
\usepackage{rotating}

\newtheorem*{hypothesis*}{Hypothesis}

\theoremstyle{definition}

\renewcommand{\thetable}{\arabic{table}}

\date{October 2024}

\title{ \textbf{Shared Hardships Strengthen Bonds: Negative Shocks, Embeddedness and Employee Retention} }

\author{Andrew Balthrop \qquad Hyunseok Jung\thanks{Balthrop: Department of Supply Chain Management, Haslam College of Business \& Baker School of Public Policy and Public Affairs, University of Tennessee, Knoxville, TN, balthrop@utk.edu, Jung: Department of Economics, Sam M. Walton College of Business, University of Arkansas, Fayetteville, AR, hj020@uark.edu.}}

\begin{document}

\maketitle
		\begin{abstract}
		\onehalfspacing 
		Jarring events inspiring reflection, known as ``shocks" in the literature, are the motive force in explaining changes in employee embeddedness and retention within the unfolding model of labor turnover. Substantial research effort has examined strategies for insulating valued employees from adverse shocks. However, this paper provides empirical evidence that unambiguously negative shocks can increase employee retention when underlying firm and employee incentives with respect to these shocks are aligned. Using survival analysis on a unique data set of 466,236 communication records and 45,873 employment spells from 21 trucking companies, we show how equipment-related shocks tend to increase the duration of employment. Equipment shocks also generate paradoxically positive sentiments that demonstrate an increase in employees' affective commitment to the firm. Our results highlight the important moderating role aligned incentives have in how shocks ultimately translate into retention. Shared hardships strengthen bonds in employment as in other areas.\\
		
\end{abstract}
\newpage
\doublespacing

\section{Introduction}

Driver turnover for trucking companies routinely exceeds 90\% per year \citep{Goodman_al2022}, costing trucking companies from \$7,894 \citep{Suzuki2007, Trick_al2021} to \$15,705 per driver \citep{Rodriguez_al1998} in 2024 dollars.  Driver turnover increases freight prices in supply chains \citep{Miller_al2020}, with increases in transport costs ultimately borne by consumers \citep{CDFJO_2023}. Higher turnover is also associated with higher accident rates \citep{MSRK_2017, Shaw_al2005, CorsiFanara1988}. Given the substantial direct and indirect costs, turnover is a perennial concern within the trucking industry \citep{ATRI_2023}, and a continuing area of research in supply chain management and logistics.

In comparison to the national average voluntary separation rate of 2.2\%, trucking industry turnover is extreme.\footnote{December 2023; Latest figures: \url{https://www.bls.gov/news.release/pdf/jolts.pdf} } These high rates of turnover provide variation that makes the trucking industry a particularly useful testbed for understanding successful worker retention strategies more generally. There are two important features of trucking, in particular, that make worker retention challenging. First, the job is highly-routinized across firms. Employees develop and depend little on firm-specific human capital so that their skills are highly substitutable between firms. In this environment, employers have little ability to offer firms-specific incentives to retain workers \citep{KCC_2021}. Second, the job is carried out remotely. While remote work, or work-from-anywhere policies have become important post-Covid,  providing valuable amenities to workers and productivity gains to firms \citep{Choudhury_al2021}, this trend has a notable downside: it reduces employee embeddedness within firms because it makes it more difficult to build meaningful relational links with co-workers \citep{Yang_al2022}. These features limit the strategies managers within the industry have for retention.

With truck drivers less tied to their job through relational linkages, they are \textit{more} sensitive to the random and often unexpected events that cause employees to re-examine their current employment relationship. These events are known in the literature as ``shocks" \citep{Lee1994, Mitchell_al2001, Lee_al2004}. Additionally, deprived of relational options for embedding employees within an organizations, managers may be tempted to treat or prevent the shock, rather than addressing the underlying job-fit characteristics, which the shock highlights. For example, recent studies have  found shocks such as pay \citep{Conroy_al2022} or schedule \citep{Bergman_al2023} variance increase turnover. A naive policy would be to defend against these shocks, such as by reducing pay frequency so that workers experience fewer of these negative events. However, while this is sometimes the only option for managers, it is the equivalent of palliative care, failing to treat the fundamental job-fit mis-alignments the shock highlights.

The primary contribution of our paper is providing important empirical evidence that unambiguously deleterious events from either the firm or employee's perspective, can result in increased retention and increased organizational commitment. We use data collected by a third-party human resources consultancy of weekly employee interviews of truck drivers for 21 trucking companies. We pair these weekly telephone interviews with employment records to follow over 45,873 employment spells. We aggregate communications records to identify 14 classes of shocks commonly faced by truck drivers, from equipment failures, and pay complaints, to issues of too little time-off.  Through statistical analysis of employment durations, we show that shocks where firm and employee incentives are aligned decrease driver turnover. In our sentiment analysis, equipment failures are followed by increased affective organizational commitment by drivers \citep{Meyer_Allen1991}.  Our empirical analyses use fixed effects and flexible model specifications to control for heterogeneity across carriers, years, months, retention programs and agents. We also consider a number of alternative specifications and re-classifications of the shocks to demonstrate the robustness of our results.

To understand the interesting pattern, we borrow the unfolding model of labor turnover \citep{Lee1994}, where shocks cause employees to re-examine their current employment relationships. From the perspective of the employer, these shocks can be ``positive'' (reducing turnover), or ``negative'' (increasing turnover). Scare quotes are necessary because the ultimate effect of shocks is moderated by the extent to which an employee is \textit{embedded} within a firm, that is, how the shock itself highlights existing job fit, relational linkages, and sacrifices \citep{Lee1994, Mitchell_al2001, Lee_al2004}. Harmful events with respect to other firm or employee objectives can increase retention. For example, an equipment failure is bad news, both for the firm and the driver. Drivers are paid piece-rate, so that inoperative equipment reduces their ability to earn. The firm, as residual claimant in this context, has similar incentives towards maintaining equipment at a high-level. When equipment fails, it highlights an important area of mutual interest alignment between the driver and the motor carrier organization, paradoxically increasing retention.

Our analysis yields a number of new and practical results. First, our extensive communications records allow us to assess the importance of a comprehensive set of work-related shocks on worker retention simultaneously. By far the most common are equipment-related shocks (33.7\%), and pay-related shocks (18\%). Whereas shocks where interests are aligned, such as those related to equipment, tend to reduce turnover, pay-related shocks result in increased turnover precisely because interests are not aligned.  
We also contribute by performing a sentiment analysis allowing us to view the shocks from two distinct theoretical lenses \citep{Philip_Craven2022}: embeddedness theory \citep{Mitchell_al2001}, and organizational commitment \citep{Meyer_Allen1991}. Shocks that highlight interest alignment have positive associated sentiments, demonstrating increased commitment. 

The remainder of the paper is organized as follows. Section 2 reviews selected job turnover research. We develop our formal hypothesis in  section 3. Section 4 discusses the dataset and empirical model. Section 5 presents results. Section 6 concludes the paper, summarizing our results, discussing research implications, and suggesting areas for future research.

\section{Literature Review}
Worker retention and turnover have been studied extensively in management \citep{Holtom_al2008}, sociology \citep{Granovetter2017}, psychology \citep{Hom2017} and economics \citep{Rogerson_al2005}. Worker turnover is costly, incurring not only direct costs in worker replacement, but can also be accompanied by productivity losses \citep{Ton2008}, reduced performance \citep{Li2022, Hancock_al2013}, product reliability concerns \citep{Moon2022} and knock-on effects through contagion \citep{Porter2021}. Understanding and reducing voluntary employee separation (hereafter, ``employee turnover"), is therefore a continuing concern. Our research contributes to two streams of research: employee turnover \citep{Hom2017} and, more specifically, to understanding turnover within the trucking industry \citep{Conroy_al2022}.
 
 \subsection{Employee Turnover}

Several meta-analyses synthesize turnover predictors, including \citet{Park_Shaw2013} and  \citet{Zimmerman_Darnold2009}. Predictors can be associated with job, training, managerial, individual, and macroeconomic characteristics. Well studied job-related turnover determinants include role clarity, job routinization,  promotion opportunities, performance and pay. The job characteristics, when combined with managerial aspects (satisfaction with leadership, influence in decision-making, team cohesion) and work-environment characteristics (stress, communication, distributive justice), determine key employee attitudes, which are strongly predictive of voluntary turnover, including overall job satisfaction, and organizational commitment \citep{Allen_al2010}.

 Nonetheless, there are a number of empirical challenges that have been difficult to overcome. Much of the turnover literature relies on exit surveys, which must then deal with issues of recall \citep{Hom2011} and selection bias \citep{Pecker_Fine2015}, and has been shown to be less useful than attitude surveys of current employees \citep{Fottler_al1995}. In particular, reports from former employees can be biased by post-decision rationalizations or overstated or made-up justifications \citep{Griffeth_Hom2001, Westaby_2005}. Moreover, causal empirical assessment requires defining a counterfactual control group \citep{MorganWinship2007}, which is impossible in exit surveys where all employees are in the ``treated" group of exiting employees. 
 
 Our dataset, based on contemporaneous communications records from weekly-periodic driver checks-ins offers a number of advantages. First, the comparatively high frequency records reduce concerns of recall bias. They also allow us to examine heterogeneous response to shocks over time that hindsight  misses. The evolving response to shocks is also theoretically important, particularly within the framework of the unfolding model of labor turnover \citep{Lee1994, LMWF1996}, in which shocks are posited to have dynamic effects. Second, in contrast to studies relying on objective administrative records, our communications records add the color of employee perceptions and allow us to employ sentiment analysis to understand employee's affective response to shocks. 

Our paper is closely related with other empirical studies examining shocks and turnover such as \citet{Conroy_al2022, Bergman_al2023, Donnelly_Quirren2006, Holtom_al2005}. However, rather than focusing on specific shocks \citep{Conroy_al2022, Bergman_al2023}, we take and integrative approach examining a broad range of shocks, including pay \citep{Conroy_al2022}, scheduling \citep{Bergman_al2023}, time off \citep{Westman_Etzion2001}, training \citep{Memon_al2016}, health \citep{Josephson_al2008}, safety \citep{Huang_al2016}, and collegiality \citep{Glambek_al2014}, as well as others. The regularity and variety of shocks illustrates two important points for human resource managers seeking to limit turnover \citep{Holtom_al2005}: first, it is impossible to insulate employees completely against shocks; second, it is not always desirable to do so, as we show that harmful shocks can unexpectedly increase retention. 
 
\subsection {Employee Turnover within the Trucking Industry}

Turnover is a particular concern in trucking, where extreme turnover rates generate extensive industry\footnote{\url{https://www.trucking.org/news-insights/truth-about-trucking-turnover}} and academic interest \citep{Miller_al2021}. 

Studies have examined trucking turnover determinants that are common to other industries, for example, worker demographics, education and union status \citep{Beilock_Capelle1990, Shaw_al1998}, pay \citep{Lemay_al1993, Phares_Balthrop2022, Conroy_al2022}, tenure \citep{Burks_al2006}, economic climate \citep{Prockl_al2017, Burks_Monaco2019, Miller_al2020}, advancement opportunity \citep{Min_Lambert2002}, and fairness \citep{Cantor_al2011}. The logistics literature has also identified factors more particular to trucking, such as home time \citep{Keller2002, Johnson_al2009, Johnson_al2011, Burks_al2023} attitude towards dispatcher \citep{Richard_al1995, Keller_Ozment1999, Morrow_al2005}, lane balance \citep{Sersland_Nataraajan2015}, safety climate \citep{Williams_al2011, Swartz_al2017}, and emotional stress and loneliness \citep{KKK2013, WTL2017}.

Much of the previous retention literature relies on cross-sectional surveys of comparatively few respondents ($<600$), with the key dependent variable being \textit{intention to leave}, e.g., \citet{Prockl_al2017, Swartz_al2017, Schulz_al2014}. This has a few drawbacks. Stated preference (\textit{intention}) can be poorly indicative of actual behaviors \citep{Chandon_al2005}, and causal assessment from cross-sectional data requires comparatively strong assumptions about the data generating process \citep{MorganWinship1999}. A separate stream of research focuses on macroeconomic aggregates and market conditions \citep{ Miller_al2021, Burks_Monaco2019, Phares_Balthrop2022, Burks_al2023}, using individual and industry-level information to understand turnover. These studies do not have the same level of intra- and cross-firm variation and lack the granularity to identify the effect of specific shocks on retention. 

This paper fills an important gap. Our dataset has significant within and cross-firm variation, unique in the literature. Additionally, its panel structure allow for us to control for unobserved time-invariant firm heterogeneity.  Indeed, the data uncovers significant cross-firm variation in retention. Weekly call intervals and the rich characterization of driver complaints allows us to address previously identified retention correlates simultaneously, tracing out the evolving effect of shocks on retention. The outcome is that we are  able to uncover paradoxical correlations that highlight the importance of interest alignment and job-fit within the trucking industry.

\section{Theory and Hypothesis}
\subsection{The Unfolding Model of Labor Turnover and Job Embeddedness}

We use the unfolding model of labor turnover \citep{Lee_al2004, LMWF1996} as our primary theoretical scaffolding for hypothesis development. This theory unifies two classes of explanations for voluntary employee turnover: external market driven reasons (``pull-related"), such as higher-wage employment options in other occupations \citep{Phares_Balthrop2022} and internal psychological-push reasons related to the employee (``push-related"), such as too little home time \citep{WTL2017}. The unfolding model draws on image theory \citep{Beach1990}, in which decisions are made through a screening process as opposed to an exhaustive evaluation of options, as under expected utility theory. Agents, awash in information and stimuli, evaluate a subset of the stimuli, determining whether the events are consistent with their objectives, or require behavioral adjustments. The unfolding model posits that a worker's decision to leave a job can be more readily explained by the arrival of new impetuses, triggering reflection and job re-evaluation, rather than by objective changes in the ecosystem of employment options. The stimulus must be sufficiently jarring to give the employee cause for reflection, an event referred to as a ``shock."

Labor turnover proceeds along four possible pathways in the unfolding model. Exit along pathways 1-3 is triggered by shocks.  Path 1 is a ``script-driven" automatic pathway, where an event occurs that triggers a planned and immediate exit. Examples include surpassing a savings target, or being assigned to a new sales territory when a worker has had prior experience with this event. Path 2 is a ``push'' driven path without a pre-planned response and without a specific job alternative in mind. An example might be being assigned to a new sales region when a worker \textit{has not} had this experience previously. Path 3 encompasses ``pull'' shocks where a specific job alternative enters consideration, as when an employee becomes aware of another job possibility. Here it is important to emphasize that it is the awareness of the job alternative, not the job alternative itself, that explains retention. In path 4, a worker decides to leave because of evolving job dissatisfaction. Path 1 exits are the most immediate. Path 2-3 exits requires time for reflection and decision, occurring over a few days to a few weeks. Path 4 exits unfold over the longest horizon of months and years \citep{mitchell2001keep}.   For our purposes, the usefulness of the unfolding model is in its dynamical treatment of shocks and the comprehensive event-centered organizing principles designed to provide tactical guidance to managers in day-to-day business operations.

Human resource managers can focus on two basic strategies for retaining employees: preventing or responding to shocks, or hardening employees against negative shocks. The latter strategy--the reason employees stay--is the complement of turnover, job embeddedness \citep{Mitchell_al2001, LBM2015}. Job embeddedness can be internal to the firm, or external, as when related to family, community, or other considerations. Dimensions of embeddedness applicable both internally and externally include links, fit and sacrifice. 

\begin{itemize}
\item \textit{Links} are people-centered and include the relationships with colleagues, mentors, and management within the organization. Employees with many links within an organization stay longer. Links can also be external to the organization, such as family considerations which keep an employee at a job, or links to industry or trade groups, and non-work friendships. 

\item \textit{Fit} regards an employee's perceived satisfaction within an organization including whether the employee is able to achieve personal goals, whether values and interests are aligned between the individual and organization, and whether the employee has the requisite skills, training and experience in their particular role. External to the organization, fit encompasses the degree to which the employee is comfortable outside of the organization, including the enjoyment of local amenities, local political and religious attitudes, and so on.

\item \textit{Sacrifice} conceptualizes how easy an individual finds it to dissolve and rebuild links and fit. Sacrifice also embodies the switching cost between jobs. Other things equal, employees are more likely to remain in their current position if leaving involves significant loss. For example, employees only partially vested in the company retirement plan will find it more costly to leave than when they become fully vested. 

\end{itemize}

Research has shown embeddedness both mediates and moderates the effect of shocks, with complex relationships with performance and organizational citizenship \citep{LBM2015, Jiang_al2012}. Shocks cause reflection on links, fit and sacrifice,  changing embeddedness (mediation). When links and fit are especially strong, shocks have a much weaker effect on the continued employment decision than otherwise (moderation). Conversely, for weakly embedded employees, comparatively mundane events can trigger employee exit. 

Job embeddedness is a latent construct whereas shocks are observable. To measure embeddedness, researchers have used measures solicited from surveys designed for the purpose \citep{mitchell20015, crossley2007development}. The concept of embeddedness has been useful in a variety of job settings, with embeddedness firmly established as an important correlate of turnover in meta-analyses \citep{ParkMin_2020, Peltokorpi_al_2015, Jiang_al2012}. As with other archival research, we do not observe embeddedness directly.  Instead, we use embeddedness theory as an organizing framework for understanding how context specific forces and shocks translate into retention through the unfolding model of labor turnover \citep{holtom2006integrating}. 

\subsection{Hypothesis Development}
\label{hyp}

As a class, truck drivers are much less embedded within a job than workers in other occupations because external embeddedness considerations rarely fix workers to a particular job. For example, a truck driver does not need to move locations to take a job with a different company, even if that company is located far away. Truckers are also less internally embedded. Drivers work remotely, making them less beholden to relational links that are important in other occupations. Additionally, skills are highly transferable across firms so that there is comparatively little sacrifice involved in changing jobs. Truck driver embeddedness relies more narrowly on considerations of job fit, resulting in less job attachment and more sensitivity to shocks.

The types of shocks that are relevant to retention are specific to the particular circumstance of the job.  Nurses are exposed to a different vector of shocks in the line of their duties than teachers or law enforcement officers. For trucking, we focus on 14 categories of shocks commonly faced by truckers, which have been highlighted by previous academic and industry research. These shocks are summarized in  Table \ref{tb1}, including categories related to Home time, Waiting, Pay, Equipment, Safety/Ethics, Training, Operations, Disrespect, Company Policy, COVID, Personal and Health. Superficially, the shocks would seemingly negatively effect retention. 

\begin{itemize}
\item Drivers work remotely away from home and family. Routes and loadings are often planned in advance to minimize empty truck miles, leaving the driver committed to an itinerary. Shocks involving home issues, such as a missed family event, an emergency requiring the driver's presence at home, or an unanticipated load pick-up that extends the driver's time away cause the driver to reflect on their current employment relationship along paths 1 and 2 and 4 in the unfolding model. Previous research has demonstrated the importance of home time to driver retention \citep{Taylor_al2010, Williams_al2011, Johnson_al2009, Johnson_al2011, Burks_al2023}.

\item Pay is the most obvious shock related to path 3 in the unfolding model, although it can also involve other paths. Pay-related issues are commonly cited within the retention literature \citep{Allen_al2010}. For tucking specifically, where drivers are paid piece-rate by mileage, pay issues can also manifest as scheduling and mileage issues that keep drivers from achieving earnings targets \citep{Williams_al2011, Prockl_al2017, Conroy_al2022}.

\item Wait time affects driver's time away from home and their realized hourly wage rate. Generally, truckers are not compensated while they wait to be loaded or unloaded, or during mandatory rest intervals, increasing uncompensated hours \citep{Belzer_2000}. Wait time also lowers the firm's effective equipment and driver utilization rates, diminishing firm earnings. Exploiting electronic logging device data, \citet{Correll_2024} finds that wait time is highly predictive of turnover. We expect wait-time to be primarily push-driven on pathway 2, but may also operate pathways 1 and 4. 

\item Equipment breakdowns and maintenance-related issues prevent safe operations and inhibit the driver's ability to earn. Trucks also provide shelter and amenities to drivers on long-haul routes, contributing significantly to ensure driver comfort. Investing in equipment is a long-known and common strategy for recruiting and retaining drivers \citep{Rakowski_al1989, Williams_al2011}. We expect equipment shocks to operate through pathways 1,2 and 4. Importantly, safe and consistent operation of equipment is also in the firm's interest.

\item Safety and ethical ethical concerns are also a common reason for voluntary turnover \citep{Williams_al2011, Swartz_al2017}. These concerns include violation of government or company rules, such as exceeding hours-of-service limits \citep{Scott_al2021}, unsafe loading or driving conditions, or equipment that violates FMCSA rules \citep{Balthrop_al2023}. Serious concerns about unsafe vehicle operation and unethical business practices likely operate strongly along path 1 in the unfolding model of labor turnover. Less serious concerns involving some amount of reflection affect retention through paths 2 and 4.

\item Training concerns include complaints related to training pay, training-related accommodations and travel, training content and duration. The effect of training itself on retention can be ambiguous depending on its portability, visibility and credibility \citep{DietzZwick_2021}, but training-related \textit{complaints} increase push-related turnover \citep{Consenza_al2012, Williams_al2011} likely along paths 1,2, and 4. 

\item Team problems include personal problems with teammate (e.g., hygiene), scheduling issues related to teaming, team-related training, and issues related to team lifestyle (e.g., cramped conditions). A good team environment provides relational links that may keep a driver at a firm, but team-related problems do the opposite. We expect team issues to be primarily push-related, operating through pathways 1,2 and 4.

\item Operations concerns, including issues such as routing, fueling, loading/unloading practices, and issues related to trailer pick-up and drop-off \citep{Johnson_al2009} can act as shocks causing re-consideration of employment along push pathways 1,2 and 4. 

\item Disrespect includes issues related to confrontational dispatchers and driver's treatment by firms' and customers' employees. Disrespect is a pervasive driver experience \citep{WTL2017}, and may contribute to push-related exit on paths 1,2 and 4.

\item Company policies related to company size, culture, customers, and rules requiring driver-facing cameras and sleep apnea machines can be a source of irritation to drivers resulting in exit along paths 1, 2, and 4.

\item Covid. The Covid-19 pandemic provided a strong impetus for exiting  \citep{sperry2022trucking}. Trucking was considered a frontline critical industry. But travel to hotspots, loss of access to supporting infrastructure (rest-stops, fueling stations, restaurants) and exposure to potential infection provided strong reason for drivers to leave along path 1 of the unfolding model. 

\item Personal issues include family issues aside from home-time concerns, retirement decisions, preference for being a leased-operator, and commute to work. We expect personal issues to result in push-driven exit along paths 1,2, and 4. 

\item Health concerns unrelated to Covid are also an important source of retention issues. Trucking is a stressful profession associated with a variety of physical and mental health concerns \citep{WTL2017, Aryal_al2023}. Health issues provide a script driven reason to exit along path 1.
\end{itemize}

We emphasize that the listed shocks are all negative events from the driver's perspective. The ultimate effect of the shock on retention, however is influenced by embeddedness, and, in particular, what the shock highlights with respect to job-fit. The effect of a shock on retention is predictable based on interest alignment between firm and employee. 

There are shocks in trucking where drivers and employer interests are aligned. Equipment maintenance is one example. Long-haul truck drivers live out of their vehicles so that keeping vehicles in good condition contributes substantially to driver welfare \citep{Williams_al2011}.  Drivers are liable for any maintenance problems that result in potential safety hazards.  Additionally, because drivers are often paid piece-rate by mileage \citep{ju2022pay}, any equipment failures that prevent vehicle operation cuts into the drivers’ pay . Drivers therefore have the incentive to immediately address and report equipment-related problems.

Correspondingly, employers have the incentive to immediately respond and take substantive action to fix equipment-related problems. First, equipment-related problems that prevent a vehicle from operating hurt firms through reduced operational efficiency and potential late-delivery penalties from shippers. Second, severe equipment problems may result in safety violations and even legal liabilities, negatively affecting a firm’s safety record and thus their ability to do business \citep{miller2022asset}.

We hypothesize, then, that equipment-related shocks underscore employee-employer interest alignment, highlighting the employees sense of of job-fit, reducing turnover. Other examples of shocks where interests are aligned are in resolving operational inefficiencies, such a driver detention, inefficient routing, and wait time. For drivers constrained by hours-of-service regulation, operational inefficiencies and wait time reduce the effective hours drivers have to cover miles \citep{Scott_al2021} resulting in reduced earnings.  Again, the motor carrier firm's incentives with respect to wait time and operation are aligned, because these inefficiencies reduce the productivity of both drivers and their equipment, reducing the firm's earnings.

Interest alignment can also result in substantive leader-member exchange leaving the employee further embedded into the motor carrier \citep{LBM2015}. When equipment, operations or wait time shocks are addressed immediately and effectively by management, such a by the dispatch of maintenance for faulty equipment, or the adjustment of itineraries to minimize driver wait time at loading docks, the effect is to leave the employee feeling that they are an important part of business operations, and that their concerns are received by management and taken seriously. Other things equal, these type of shocks make it more likely for employees to stay.

Driver and employer interests are not aligned for most of the listed shocks. Pay and home time provide important counter-examples \citep{Burks_al2023}. Piece-rate pay per mile implies that any pay increase for drivers reduces the residual for the firm. Pay increases have not been shown to increase driver productivity in this industry \citep[see, for example,][on income targeting]{Belzer_Sedo2018} so there are no output increases--either. Thus, pay-rates are a pure cost which the firm has incentive to minimize. Similarly, more home-time for drivers, means less capacity for the firm and lowered capital utilization when drivers take vehicles home \citep{Burks_al2023}.  Conversations between employers and employees about pay and home time are thus more likely to be acrimonious and less likely to result in fruitful resolution. The resultant interest divergence reduces embeddedness, increasing turnover.

Driver and Employer interests are misaligned for the majority of shocks we study, including home-time, pay, training, teams, disrespect, company policy, Covid, personal matters, and health. In none of these cases do employee issues attach to the fundamental business operations determining firm profitability, except through the effect on employee retention.

The safety/ethics category of shocks present an interesting edge case in terms of interest alignment. For severe safety issues potentially resulting in a crash, interests are certainly aligned, not least because of the potential legal liabilities for the firm, and injuries to drivers. However, our data sample is made up of comparatively safe firms, so we do not observe many of these types of issues. Instead, reported ethical transgressions, such as dishonesty in recruiting, are comparatively more frequent and clearly misaligned with respect to employee and employer interests. We therefore provisionally classify safety/ethics as misaligned.

Taken together, we have:

\begin{hypothesis*}
Negative driver experiences which highlight job-fit through employee-firm interest alignment increase employee retention. Negative experiences where interests are not aligned decrease retention.
\end{hypothesis*}

\section{Data and Empirical Model}
\subsection{Data} \label{data}

Our empirical analysis uses communication records from a human resource outsourcing company specializing in retention programs for trucking companies. Primarily designed for new drivers, motor carriers typically apply these program for all new hires from training and orientation to three or six months after, depending on the program.\footnote{Program details are determined by carriers. Our hazard models use fixed effects and flexible hazard function to control for heterogeneity across carriers, call center agents, and programs. See section \ref{coxph} for further details.} The core of the programs consists of weekly (or longer interval) phone calls to truck drivers to identify and record their ``issues''-- ``shocks'' in the terminology of the unfolding model of turnover-- and then pass this information to their customers (i.e., trucking companies) to help them to be proactive in driver retention. Specifically, during the program, an agent of the human resource company records a brief summary of the driver conversation after each call, tagging it with proprietary classification codes for the issues identified during the conversation 
(hereafter issue codes). Some examples of the communication notes are below:
\begin{itemize}
    \item \textit{Good afternoon Mr. .... All is well sir the miles are rolling in well the road has been good thank you. Thank you for checking in. Be well!}
    \item \textit{Not great. She decided to quit as she was messed over on her apprenticeship hours and all that. She hated that they sent her all the way to NY for her training.}
\end{itemize}

The classification system includes more than two hundred issue codes that have nested structure allowing for coarse, intermediate and granular characterization of issues.\footnote{Due to an non-disclosure agreement with the human resource company, the full list of issue codes and their definitions cannot be shared.} For example, a non-competitive pay rate is classified as ``Pay: Rate: Non-competitive," while an unclear mileage rate is classified as ``Pay: Rate: Confusing/Unclear."  In many cases, coarse issue codes share common cross-cutting intermediate categories, as when Covid issues show up as sub-categories in Pay, Training, Personal, etc.,. These cross-cutting categories allow masking of proprietary trade secrets by re-organizing the data without losing important issue-code variation.  We re-classify primitive issue codes into the fourteen major categories of section \ref{hyp}: Home-time, Pay, Equipment, Safety/Ethics, Training, Team, Operations, Disrespect, Company, Covid, Waiting, Personal, Health, Others. In our main analysis, we further aggregate the categories into two grand classifications: ``Aligned," where employee-employer interests are aligned, including Equipment, Operations and Waiting, and ``Misaligned," where their interests are opposed, which includes all other categories.

We merge these communication records with driver employment data to create survival-time data suitable for duration analysis. The employment data includes hire and termination dates, from which we calculate employment duration (in weeks) and termination status at the time of each communication record. For drivers whose employment was not terminated during the program, no termination date is observed and their employment spells are treated as censored (i.e., right-censoring). To create a more homogeneous sample, we exclude non-voluntary termination cases (i.e., disqualified quit) and employment spells longer than 100 weeks.\footnote{Our results in the next sections  are robust to this sample selection.} The data is then merged with employer data such as carrier-level scores related to unsafe driving, hours-of-service compliance, vehicle maintenance, and misuse of controlled substances/alcohol from Safety Measurement System (SMS) of the Federal Motor Carrier Safety Administration.

The duration and termination status variables, along with the communication records, are the key variables in our duration analysis. 
Our data is multi-record-per-subject survival data, where individual drivers are observed multiple times with different communication records over their employment spells. This provides our analysis additional strength because it allows for time-varying regressors that can account for the entire history of covariates. This feature is particularly useful for analyzing the effects of lagged variables.

Our final sample includes 466,236 communication records (= total observations) of 45,873 employment spells from 21 carriers over 2019 - 2022. Most spells are from large carriers, with the largest carrier accounting for 60\% and other seven carriers accounting for 32.6\% of the sample.\footnote{Employment concentration at larger firms is typical in this industry. While the majority of firms are small, the power-law distribution of firms size dictates that the majority of workers work for larger firms \citep{Balthrop2021}.} Later years includes more observations: 16,882 (2019), 71,839 (2020), 158,221 (2021), 219,294 (2022).\footnote{These are the numbers of communication records from the spells initiated in a given year.} The sample includes four different retention programs, four call-intervals, and fifteen agents, where two programs account for 97.8\% of the sample and most observations are based on weekly or bi-weekly phone calls (99.8\%). Our duration analysis controls for heterogeneity across carriers, years, months, programs, call-intervals and agents using fixed effects and flexible hazard functions.\footnote{In practice, the retention programs and call-intervals are determined by carriers, and in the data, program type and call-interval do not vary much within carriers (i.e., each carrier applies the same program and call-interval to all their drivers). This indicates that controlling for carrier-level heterogeneity is particularly important and an effective way to control for overall heterogeneity in our data.}

The distribution of the issue codes under the fourteen major categories is provided in Table \ref{tb1} below. The most frequent issue was Equipment, followed by Pay, Disrespect and Safety/Ethics. In terms of the ``Aligned" (i.e., Equipment, Waiting and Operations) and ``Misaligned" (i.e., the rest) categories, about 47\% of the codes issued belong to the ``Aligned" category.

\bigskip \centerline{\lbrack === Table \ref{tb1} here ===]}\bigskip

\begin{figure}[h!]
	\caption{Kaplan–Meier Survival Curves}
		 \label{fg1} 
	\includegraphics[width=1\linewidth, trim={0cm 0cm 0cm 0cm},clip]{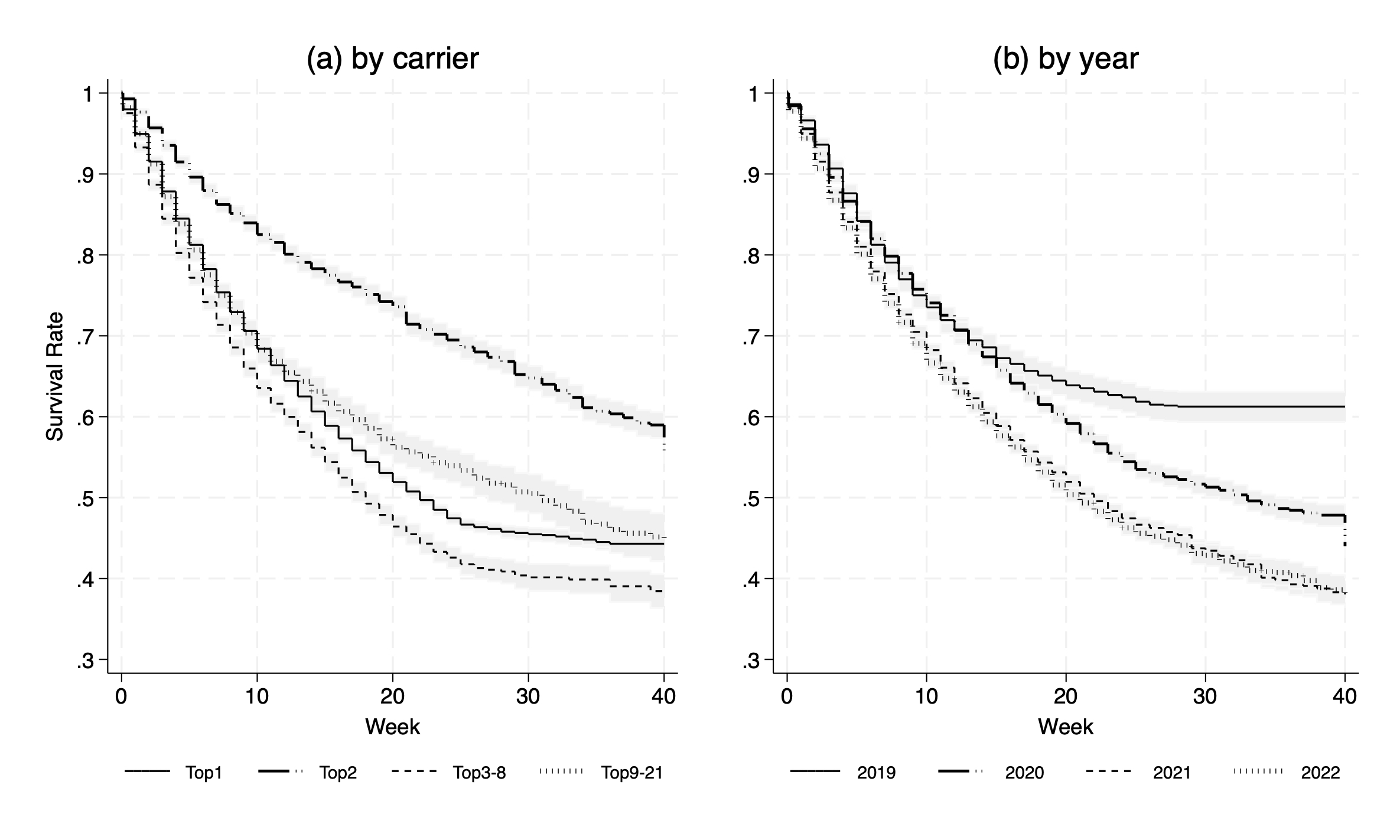}
\end{figure}

\begin{figure}[h!]
	\caption{Survival and Hazard Curves for Driver Groups with Different Types of Issues}
		 \label{fg2} 
	\includegraphics[width=1\linewidth, trim={0cm 0cm 0cm 0cm},clip]{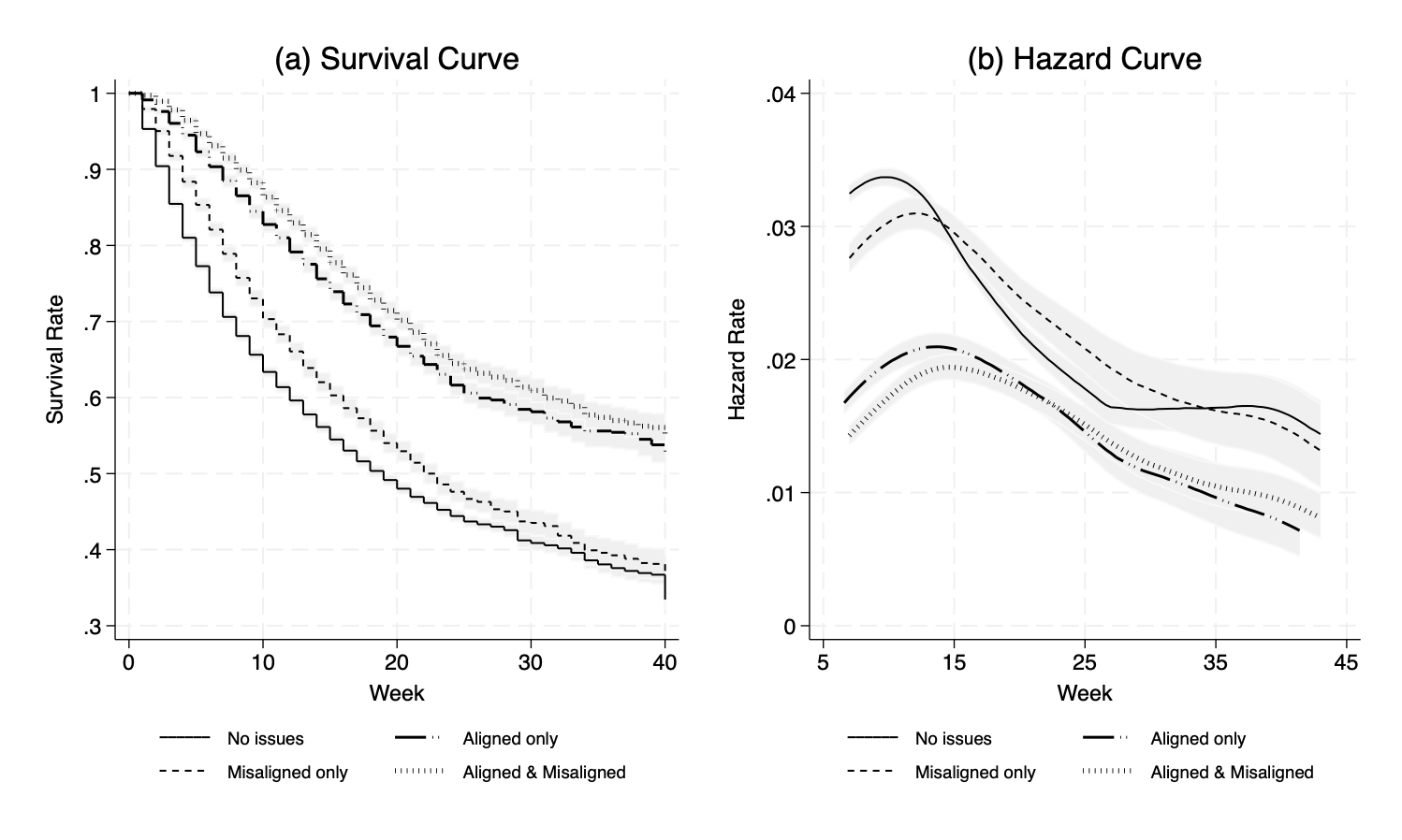}
\end{figure}

Next, we provide Kaplan-Meier (KM) survival curves and smoothed hazard estimates to summarize the important features in the data. The survival and hazard curves in Figures \ref{fg1} and \ref{fg2} plot survival rates, i.e., the proportion of employment spells that did not terminate after a given number of weeks, and hazard rates, i.e., the proportion of employment spells that terminated at a given week among those not yet terminated by that time, for each employment week. The shaded areas represent 95\% confidence intervals.

\bigskip \centerline{\lbrack === Figure \ref{fg1} here ===]}\bigskip 

Figure \ref{fg1} includes survival curves by carrier and year, highlighting important heterogeneity in the data. The carriers are ranked in terms of the number of employment spells and their rankings are denoted by ``Top-." The survival curves demonstrate the high turnover rate in trucking industry. Top1 and Top3-8 carriers' survival rates are below 0.5 after 24  weeks, indicating that more than 50\% of the drivers in these carriers quit their jobs within six months. Significant heterogeneity across carriers is observed. The survival rate of Top2 carrier is significantly higher than the other carriers, with the gap widening as the employment duration increases. The heterogeneity across years is also interesting. The survival curve of year 2019 flattens around fifteenth week, implying that driver turnover significantly decreases after the period. However, after 2020, turnover worsens, even for long-duration employees,  providing evidence consistent with the trend of industry-wide employment re-shuffling in the wake of the Covid pandemic \citep{phares2023state}.

\bigskip \centerline{\lbrack === Figure \ref{fg2} here ===]}\bigskip 

Figure \ref{fg2} presents survival and hazard curves for four different driver groups: i) the group of drivers experiencing no issues throughout their spells (26,777 spells), ii) the group experiencing Aligned issues only (6,402 spells), iii) the group experiencing Misaligned issues only (6,715 spells), and iv) the group experiencing both Aligned and Misaligned issues (5,979 spells).\footnote{The hazard curves are smoothed by a kernel method in \citet{KM2003}.} Note that the levels and curvatures of the plots do not directly represent the effects of the different types of issues since they do not account for the timing of the issues and heterogeneity across drivers. We account for these features in our duration models in the next section when calculating the marginal effects of the issues. 

There is a significant difference in survival and hazard rates between groups experiencing Aligned issues and those not: groups experiencing Aligned issues exhibit  higher survival rates and lower hazard rates over the whole employment period. Looking at the hazard curves, the discrepancy is pronounced in the early period, leading to a significant gap in survival rates in later periods. In contrast, the differences in survival and hazard rates between ``Misaligned only" and ``No issues" is quite small; similarly, ``Aligned \& Misaligned" closely tracks ``Aligned only." 

\subsection{Empirical Model: Cox Proportional Hazard Model}
\label{coxph}
We use the Cox Proportional Hazard (PH) model to analyze the effects of issue types on driver turnover. The Cox PH model specifies the hazard function, i.e., the probability that an employment spell is terminated at $t$ given that it has not been terminated before $t$,\footnote{More precisely, in case of discrete time, the hazard function without covariates is defined as $\lambda(t) = \Pr[T = t | T \geq t]$ (= conditional failure probability), where $T$ denotes the duration random variable, whereas in case of continuous time, it is defined as $\lambda(t) = \lim_{h \to 0} \Pr[ t \leq T \leq t + h | T \geq t]/h$ (= instantaneous failure rate), which is technically not a probability, but in practice, it is loosely interpreted as ``instantaneous" probability.} as 
\begin{equation}\label{eq1}
    \lambda(t | \mathbf{x}) = \lambda_0(t) \exp\left(\mathbf{x}^{'}\beta\right),
\end{equation}
where $\lambda_0(t)$, called the baseline hazard, is a function of $t$ (duration) alone and $\exp\left(\mathbf{x}^{'}\beta\right)$ is a function of $\mathbf{x}$ (a vector of covariates) alone, so the two functions separate the hazard rate into two portions that varies by $t$ and that varies by $\mathbf{x}$. The separability is the key assumption in the Cox PH model and allows the estimation of $\beta$ without specifying the baseline hazard in the partial likelihood estimation approach proposed by \cite{Cox1972,Cox1975}. The baseline hazard function can be recovered after estimating $\beta$ using a method similar to the KM estimator. In duration models, the hazard function, which fully characterizes the distribution of duration, is the main building block that models the associations between duration and covariates of interest.

Our covariate set $\mathbf{x}$ includes the dummies for each issue category. To capture the dynamic effects within the framework of the unfoloding model of turnover, we also include their lags up to four periods.\footnote{The maximum number of lags was chosen, in a preliminary analysis, by running the duration model with the Least Absolute Shrinkage and Selection Operator \citep[LASSO,][]{lasso}, where lags beyond four periods were removed by the LASSO for most categories.} When reporting the results in the next section, we sum the coefficient estimates for the original dummy and its four lags and then divide it by five to compute the average effect.\footnote{To calculate the standard errors of these average effects, the delta method is used.} We also report hazard ratios, which calculate the impact of an issue on the hazard rate relative to the baseline hazard. Due to the parametric form of the Cox PH model and that issue covariates are all dummies, the hazard ratios are computed as $\exp(\beta) - 1$.\footnote{Note that the hazard ratio of a dummy $x$ in the Cox PH model is $[\lambda(t|x=1) - \lambda(t|x=0)]/\lambda_0(t) = \exp(\beta) - 1$.}

We use fixed effects and flexible hazard functions to control for heterogeneity across carriers, years, months, programs, call-intervals and agents. In all specifications, we include fixed effects for year, month, retention program, call-interval and agent. Also, we include a variable that quantifies the non-responsiveness of each driver to the human resource company's calls, ``Nonresponsive", which is the proportion of calls tagged with ``voice mail," ``no answer," etc. Responsiveness is an important \textit{revealed} characteristic that can be correlated with driver heterogeneity. The Nonresponsive variable is quite uniformly distributed across drivers, with quartiles at 0.20, 0.42, and 0.64, implying that the variable is not concentrated within a subset of individuals. To draw robust conclusions,  standard errors are always clustered at the spell level. 

Since much of the retention program's details (length, call-interval, etc) are determined by carriers, and carrier characteristics are arguably the most important factors in driver's quit decision, we employ three different approaches to control for heterogeneity across carriers: fixed effects, carrier-level variables from the SMS, and carrier-specific baseline hazard functions. In the result tables in the next sections, Model 1 and 2 includes carrier fixed effects and carrier-level variables, respectively, and Model 3 allows for  carrier-specific baseline hazard functions.\footnote{The partial likelihood estimation of the Cox PH model calculates the likelihood of each (observed) termination event by the hazard rate of the spell terminated at the event divided by the sum of the hazard rates of the terminated and non-terminated (i.e., censored) spells existing at the event. The division in the calculation cancels out the baseline hazard function due to the specification in (\ref{eq1}), which allows us to estimate $\beta$ without specifying the baseline hazard function. When group-specific baseline hazard functions are allowed, the ``partial" likelihoods are calculated by group.}

\section{Results}
\subsection{Effects of Issue Types on Driver Turnover}

To examine our hypothesis,  Table \ref{tb2} reports the coefficient estimates for Aligned and Misaligned issues in the Cox PH model.\footnote{Remember that the estimate for each issue type is the average of the estimates for its original issue dummy and four lags, so it includes the dynamic effects.} As defined in Section \ref{data}, Aligned, where employee-employer interests are aligned, includes Equipment, Operations and Waiting, and Misaligned, where their interests are opposed, includes all other categories. Coefficients estimates are stable across specifications.

\bigskip \centerline{\lbrack === Table \ref{tb2} here ===]}\bigskip 

A stark contrast between the two types of issues is observed: all other things being equal, Aligned issues on average decrease driver turnover, whereas Misaligned issues increase it. ``h.r." are hazard ratios, which helps the reader quickly appreciate the magnitude of the impact. For example, in Model 1, the hazard ratio of Aligned is $-0.23$, indicates that experiencing a shock that highlights aligned interests decreases the hazard (i.e., instantaneous probability of leaving the job) by 23\% over the baseline hazard. On the contrary, the hazard ratio for Misaligned issues is $0.40$, indicates having an issue like Home-time, Pay, etc., increases the hazard by nearly 40\%. The somewhat counter-intuitive result here is that Aligned issues lower driver turnover, even though they are still negative events that drivers experience. For example, a driver that experiences problems with having to wait at a loading dock is \textit{more} likely to remain at the motor carrier than a driver that experiences no wait-time shock.

We argue that this is because the Aligned issues provide opportunities for both drivers and employers to communicate with positive messages and confirm their mutual interests, thus increasing the employee's embeddedness. When the driver experiences operational difficulty, equipment failure, or extended wait times, these shocks also harm the firm by directly reducing worker and equipment utilization. A worker detained at a loading dock, or broken down on the roadside, is not transporting freight. In these cases the motor carrier as an organization has incentive to sympathize with workers, which may result in strengthened relational linkages between frontline workers (drivers) and their support teams. Beyond words, the firm is also likely to take action to remedy the problem. Whether through words (which we observe) or actions (unobserved) these Aligned shocks highlight the firm and workers' common purpose, while potentially revealing the firm's support infrastructure available to the worker.

We further disaggregate these results in Table \ref{tb3}, which reports the coefficient estimates for fourteen major issue types in the Cox PH model.\footnote{More detailed results including the estimates of individual lag terms are available upon request.} 

\bigskip \centerline{\lbrack === Table \ref{tb3} here ===]}\bigskip 

As with our main results, coefficient estimates in Table \ref{tb3} are not sensitive to specification, whether carrier fixed effects (Model 1), carrier-level variables (Model 2) or carrier-specific baseline hazards (Model 3) are included. Categories most strongly associated with increased turnover include Covid and Personal issues. While Covid is a strong indicator of potential exit, it is a comparatively rare issue (only 224 instances), even though our sample period covers the onset of the pandemic. Personal issues are more common, and have stronger effects on turnover.  This category encompasses issues related to family, which supports previous research establishing loneliness and family disruption as one of the most important reasons for truck driver turnover \citep{WTL2017}.  Even the smaller (but positive) coefficients have managerially meaningful effects on hazard:
ethics complaints increase the baseline hazard rate by 23\% (Model 1). In contrast, the Aligned issues, including Equipment, Operations, and Waiting, all result in strongly reduced hazard rates. The coefficient on Nonresponsive variable (i.e., proportion of calls not answered) is reported in Table \ref{tb2}, which is, as expected, positive and significant, with estimates ranging from 0.63 to 0.67 across models, implying drivers less responsive to the retention programs exhibit higher turnover rates.

\subsection{Robustness Check}

A potential concern, however, is that there may be unobserved driver types with different hazard rates and these types are correlated with the kinds of issues they report. For example, there may be ``attached" and ``detached" drivers with low and high turnover rates, and the former report aligned issues more frequently than the latter, simply because they are more attached to their company. Then, the signs of the coefficients on Aligned and Misaligned issues are simply a reflection of the hazard rate difference between the two types, not the effects of the two different types of issues. Said differently, it is possible to imagine a driver experiencing an equipment issue but not reporting it precisely because the driver expects the issue will go unresolved. Such a disillusioned driver can be expected to have a higher turnover rate; importantly, this type of detached driver will fall into the base-case ``no issue" category, biasing base-case turnover rates upward. The Nonresponsive variable in our models accounts for the issue in part, but this section explores various approaches to further examine the robustness of our results. We pursue three strategies: i) controlling for additional covariates, ii) focusing on a sub-sample of responsive drivers, and iii) examining for differences between employee drivers and owner-operators.

First, we include ``issue frequency" variables into our models and report the results in Table \ref{tb4}.  Issue frequencies are calculated for Aligned and Misaligned issues, each frequency being the number of each driver's Aligned or Misaligned issues reported throughout their sample period divided by the period length (hereafter Aligned\_f and Misaligned\_f). To account for potential non-linearity, we also include their squared terms. The frequency variables essentially allow us to compare outcomes among those who exhibit similar frequencies of Aligned or Misaligned issues throughout their employment spells.

After inclusion of the variables, the signs and significance of Aligned and Misaligned variables remain the same. However, the two coefficients exhibit somewhat different patterns. The coefficient on Misaligned issues decreases significantly (about 35\%), while the coefficient on Aligned issues largely remains the same. The result can be interpreted as follows. Recall that our model includes dummies for Aligned and Misaligned issues and their lags up to four periods (as stated in Footnote 11, we used the LASSO to determine the optimal number of lags). The coefficient estimate for Aligned or Misaligned issues reported in this paper is the average of the estimates for its original dummy and four lags. Therefore, the coefficient estimate roughly represents the likelihood of employment termination ``within" five weeks after Aligned or Misaligned issues. In this setup, the frequency variables are likely to capture turnovers beyond five weeks. Therefore, the decrease in the coefficient on Misaligned indicates that there were many cases where drivers did not leave the company within five weeks after Misaligned issues but eventually left the company (without any issues reported). The coefficient on Aligned issues remains the same, implying there is no such pattern for Aligned issues.\footnote{As discussed below, the coefficients on Aligned\_f and (Aligned\_f)$^2$ in Table \ref{tb4} are opposite (negative and positive), implying a nonlinear effect of Aligned\_f, which makes the overall effect to be close to zero.}

These turnovers may be seen as a long-term effect of Misaligned issues but can also be seen as a result of ``persistent" driver heterogeneity that might have affected both issue reporting behavior and turnover. Indeed, this approach is similar to an econometric treatment popular in practice for accounting for individual heterogeneity: the correlated random effects (CRE) approach \cite[e.g.,][]{W2019}. The CRE approach specifies individual heterogeneity as a function of the time-averages of the covariates in the model and includes these additional variables into the model to control for the heterogeneity. The inclusion of frequency variables can therefore be viewed as an application of the CRE approach, with the results in Table \ref{tb4} demonstrating the robustness of our conclusion to individual heterogeneity.

The coefficients on the frequency variables and their squared terms deserve close examination. The coefficient on Aligned\_f is negative but the coefficient on its squared term (i.e., (Aligned\_f)$^2$) is positive, implying a non-linear relationship between the number of Aligned issues and turnover rate. For those with relatively low Aligned issue frequencies, experiencing an additional Aligned issue decreases turnover, but when the frequencies are high, it increases turnover. Drivers reporting issues very frequently may be seen as ``detached" drivers, regardless of issue type, and the squared term captures their inherently high hazard rates. The coefficients on Misaligned\_f and its squared term are all positive, indicating higher frequency is always associated with higher turnover for Misaligned issues.


As a second strategy, we focus on the sub-population of observed communicators. That is, rather than making comparison to a reference group that may under-report issues, we make comparisons between groups that report \textit{multiple} issues.   First, we dis-aggregate the Aligned category into ``Broken-equipment" and ``Other aligned," where ``Broken-equipment" includes issues related to ``broken/missing components" only, which is arguably less subject to the under-reporting problem. Second, we re-classify the issues into new categories that account for joint-distribution of Aligned and Misaligned issues: Broken-equipment only (B-only), Other aligned only (OA-only), Misaligned only (MA-only) and their combinations.\footnote{Note that a communication record can be tagged with multiple codes in different categories. Communication records counted multiple times for the fourteen categories in Table \ref{tb1} can be counted fewer times for the new categories in Table \ref{tb5}, resulting in different total numbers in the tables.} 

We note that, if other covariates are ignored, the coefficient on Aligned in Table \ref{tb2} is calculated by a combination of two hazard rate differences: a hazard rate difference between the cases experiencing Aligned issues only and the cases experiencing no issues, and another hazard rate difference between the cases experiencing both Aligned and Misaligned issues and the cases experiencing Misaligned issues only. Roughly speaking, the first difference compares drivers reporting Aligned issues only and drivers who were silent, whereas the second difference compares drivers who report \textit{at least one} of the two types of issues. Because the second comparison is among ``communicators," it is arguably less subject to the under-reporting issue. The new categories allow us to identify these dis-aggregated effects.

\bigskip \centerline{\lbrack === Table \ref{tb5} here ===]}\bigskip 

Table \ref{tb5} includes the distribution of the issues under the new categories, and Table \ref{tb6} reports the coefficient estimates for the new categories. The coefficient on ``Broken \& Misaligned" subtracted by the coefficient on ``Misaligned" calculates the effect of Aligned issues that is less subject to the under-reporting issue as argued above, which is -0.127 in Model 1 (i.e., specification including carrier fixed effects), with standard error 0.078, t-statistic -1.63 and p-value 0.10.\footnote{The standard error is calculated using the delta method. The results from Model 2 and 3 specifications are qualitatively the same.} The p-value is 10\%, but the result overall indicates that relative to the population experiencing only Misaligned shocks, a driver experiencing \textit{both} an Aligned and Misaligned shock has a lower hazard rate. The striking result that two negative shocks are better than one shock provides strong support for our hypothesis, at least in this sub-population.

\bigskip \centerline{\lbrack === Table \ref{tb6} here ===]}\bigskip 

Third, we examine how the results change when we account for equipment ownership status. Carriers may own their tractors or lease them by contracting with owner-operators. Owner-operators are responsible for the maintenance of their own tractors; therefore, we can expect smaller effects of equipment-related shocks in this sub-population. Unfortunately, our data doesn't include information on individual-level equipment ownership, so we use carrier-level lease rates for this exercise. We calculate lease rates for each carrier by the proportion of leased tractors,\footnote{The average lease rate for 21 carriers over the years 2019-2021 is 21.02\%.} using data from the FMCSA Census. We select two sub-samples containing spells from the carriers with the average lease rate higher than 40\% over the years 2019-2021 (``Lease\% $>$ 40") and from those with the lease rate lower than 5\% (``Lease\% $<$ 5"). Each sub-sample accounts for about 10\% of the whole sample.\footnote{We use these cutoffs to select two sub-samples that are comparable in terms of size and reasonably large to get reliable results.}

Table \ref{tb7} report the results from the two sub-samples using the Model 3 specification in Table \ref{tb6}. The results from other specifications are qualitatively the same. We also report the distribution of the issue codes in each sub-sample in Table \ref{tb8}. The sub-sample ``Lease\% $>$ 40" contains fewer Broken-equipment only shocks and more Misaligned shocks than ``Lease\% $<$ 5", but the overall distribution of the issues is similar across the two sub-populations, enhancing the comparability of the results in Table \ref{tb7} across the two samples.

\bigskip \centerline{\lbrack === Table \ref{tb7} here ===]}\bigskip 

\bigskip \centerline{\lbrack === Table \ref{tb8} here ===]}\bigskip 

For Lease\% $>$ 40 (i.e., the sample including many owner-operators), most of the coefficients related to equipment shocks are now insignificant. It appears that only Misaligned and Other aligned \& Misaligned are significant predictors for driver turnover in this sample. This aligns with our prediction that if the equipment-related coefficients indeed capture the effects of equipment shocks, not only reflecting driver heterogeneity or inherent hazard rate differences between driver types, the effects will be smaller when drivers own and maintain their equipment. The results from Lease\% $>$ 40 are similar to those in Table \ref{tb6}.

\subsection{Potential Mechanism: Sentiment Change after Issues}

To explain the potential mechanisms for the turnover rate changes observed in the previous section, we employ sentiment analysis to examine drivers' affective response. We use the VADER (Valence Aware Dictionary and sEntiment Reasoner) algorithm to compute sentiment scores, ranging -1 to 1 (the larger, more positive), for each communication note in the data. The VADER algorithm is a popular lexicon and rule-based sentiment analysis tool, which combines a sentiment lexicon--a list of words labeled as either positive or negative according to their semantic orientation--with certain rules (e.g., how to handle negation, modifier, etc) to compute sentiment scores. Compared to other sentiment analysis tools (e.g.,TextBlob), the VADER is specifically attuned to sentiments expressed in microblog-like contexts (e.g., social media) \citep{HG2014}, which is appropriate for our analysis given the form and style of the communication notes in our data.\footnote{For sentiment scores, we use the compound score in the VADER, which combines positive, neutral and negative scores calculated in the algorithm.}

\begin{figure}[h!]
	\caption{Sentiment Change after Driver Issues}
		 \label{fg3} 
	\includegraphics[width=1\linewidth, trim={0cm 0cm 0cm 0cm},clip]{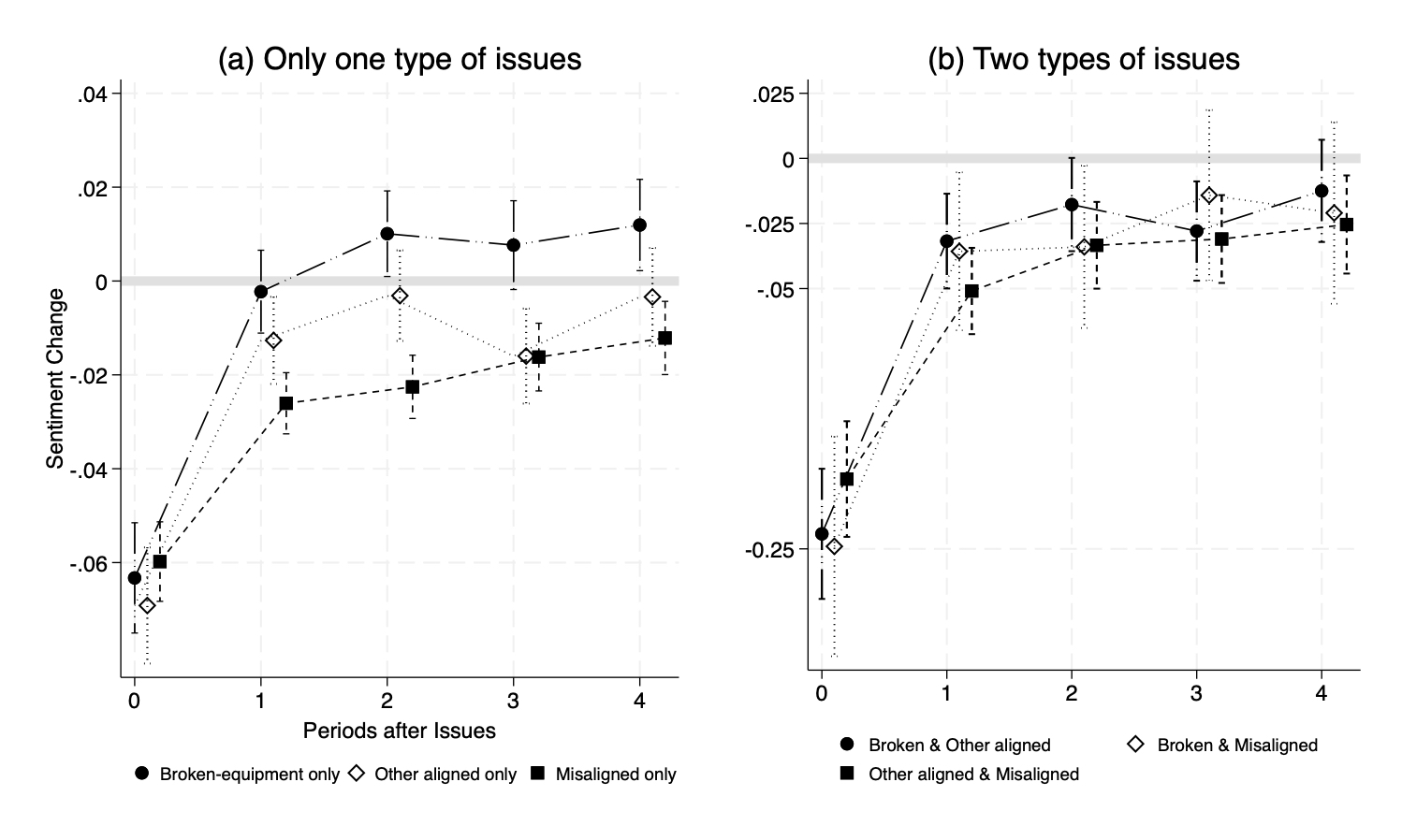}
\end{figure}

We regress the sentiment scores on the same set of regressors used in the Cox PH models in Table \ref{tb6}, and plot the coefficient estimates and their confidence intervals, including the lag terms, in Figure \ref{fg3}.\footnote{The full set of results is contained in Table \ref{tb-a2}, Appendix \ref{ap2}. Note that the scale of y-axis in the second panel of Figure \ref{fg3} is adjusted to show the changes in coefficients more clearly.} Driver issues all have a negative effect on the sentiment of conversations right after the issues. Because it is a common practice to classify sentences as positive, neutral or negative if the VADER sentiment score is greater than 0.05, between 0.05 and -0,05, or less than -0.05, respectively,\footnote{ \url{https://github.com/cjhutto/vaderSentiment\#about-the-scoring}} the sizes of the coefficients at period = 0 (i.e., the period issues were identified) implies a substantial negative effect of driver issues on conversation sentiment.

However, the pattern of how the negative sentiment changes over time is significantly different across issue types. In case of ``Broken-equipment only," the negative sentiment turns to positive after one period and the positiveness continues. On the other hand, the negative sentiment on ``Misaligned only" remains negative. The coefficients on the combinations of the three issue types are less significant due to the smaller number of such cases, but a similar pattern is observed in the second panel of Figure \ref{fg3}: when the case is more associated with ``Aligned" issues, the negative effect tends to disappear faster. These observed relationships are suggestive evidence that companies take remedial action in the event of broken equipment. The lagging positive sentiments are then the response to such actions.


\section{Discussion}

Employees experience numerous and varied shocks which cause them to reflect upon and re-evaluate their current employment relationship. This paper catalogs self-reported shocks from a sample of over 40,000 truckers, which we categorize into Aligned and Misaligned categories. The most commonly reported events are problems related to equipment (Aligned) followed by pay complaints (Misaligned). The surprising result is that Aligned shocks, other things equal, result in longer job spells. 

This finding is consistent with our hypothesis and aligns well with the framework of the unfolding model of labor turnover \citep{Lee1994}. Shocks are not direct determinants of retention, instead they highlight specific aspects of the current employment relationship in terms of links, fit, and sacrifice.  Aligned shocks have the ultimate effect of underscoring job-fit compatibility between employee and employer. Motor carriers have the incentive to address Aligned issues immediately, leaving the employee feeling listened to and supported, thereby increasing embeddedness and ultimately retention. Indeed, sentiment analysis demonstrates positive employee feelings directly following equipment-related shocks. Misaligned shocks have the opposite effect. Rather than highlighting job fit through interest alignment, these complaints underscore fundamental interest divergence at the current job. 

The primary threat to this interpretation is self-reporting bias \citep{scott2021consequences}, for instance, if unobserved employee attachment is correlated with observed communications. If employees that are particularly attached to an organization are more likely report problems, then communication is the outcome of employee attachment, rather than the cause. Our extensive data set--the largest and richest panel yet used in trucking retention research-- allows us to address these concern using three strategies: i) we control for non-response and complaint frequencies, ii) we restrict analysis to the sub-sample of observed responders, and, iii) we examine whether our results change by equipment ownership status in a theoretically consistent way. The main result stands in each case.  Negative shocks increase retention when they highlight job-fit through employee-firm interest alignment. 

\subsection{Theoretical Implications}
Our results have implications to retention theory generally, but also provide context-specific implications to retention research within the trucking sector. First, our results highlight the usefulness of the unfolding model of turnover. Our  paper finds context-specific shocks to be resoundingly predictive of employment duration, consistent with the unfolding model. It would be difficult to model and understand these correlations within the framework of competing theories, viz., expected utility theory. 

Second, our research highlights important boundary conditions within the unfolding model of labor turnover on how shocks translate into retention. When negative shocks underscore job-fit, the duration of employment increases. In the trucking industry context, it is equipment, operations, and wait-time shocks.  
At the boundary of retention lies the complementary theory of job embeddedness, which rests on job-fit, links and sacrifice.   Given our findings on how job-fit influences the ultimate effect of shocks, it is reasonable to suspect that considerations of links and sacrifice may have similar effects, at least in other industries where these concerns are more significant. 

These results also have intriguing links to an analogous stream of defensive marketing research showing how businesses are better served by devoting resources towards addressing existing customer complaints, than on offensive marketing campaigns designed to draw in new customers \citep{fornell1987defensive}. For example, by addressing service failures expeditiously, business can increase customer loyalty \citep{tax1998customer}. Similarly, in high turnover-industries, employers may be better-served through defensive expenditures on current employees rather than devoting more effort to recruitment. Our data is produced by exactly this sort of defensive investment. A key point for theory to address given sophisticated retention efforts, then, is the role of employers in ``voluntary" exits, which are generally ignored in the psychology literature: even these exits are to some extent joint decisions between employee and firm, potentially revealing a firm's exhaustion with an employee.

\subsection{Guidance for Practitioners}

While pay is one of the most common complaints for truck drivers, its effect on retention is weak in comparison with home time, company and personal issues.  While home time has been previously highlighted \citep{Keller2002, Johnson_al2009, Johnson_al2011}, the company effects have not. This category includes policies with regards to pets, driver-facing cameras, and sleep apnea machines; however, the most common complaint in the category is firm policies that do not allow a driver to take the tractor home at the end of a shift.  Given that drivers' homes, equipment yards and load locations can be separated by hundreds of miles, it is easy to understand the importance of this amenity. Managers may be willing to accept lower equipment utilization rates in favor of increased retention. Within the ``personal" category the most important issue for drivers is being away from family. Offering more amenities in terms of family support is another important measure managers can focus on to increase retention. 

\begin{figure}[h!]
	\caption{Frequency and Impact of Driver Issues}
		 \label{fg4} 
	\includegraphics[width=1\linewidth, trim={0cm 0cm 0cm 0cm},clip]{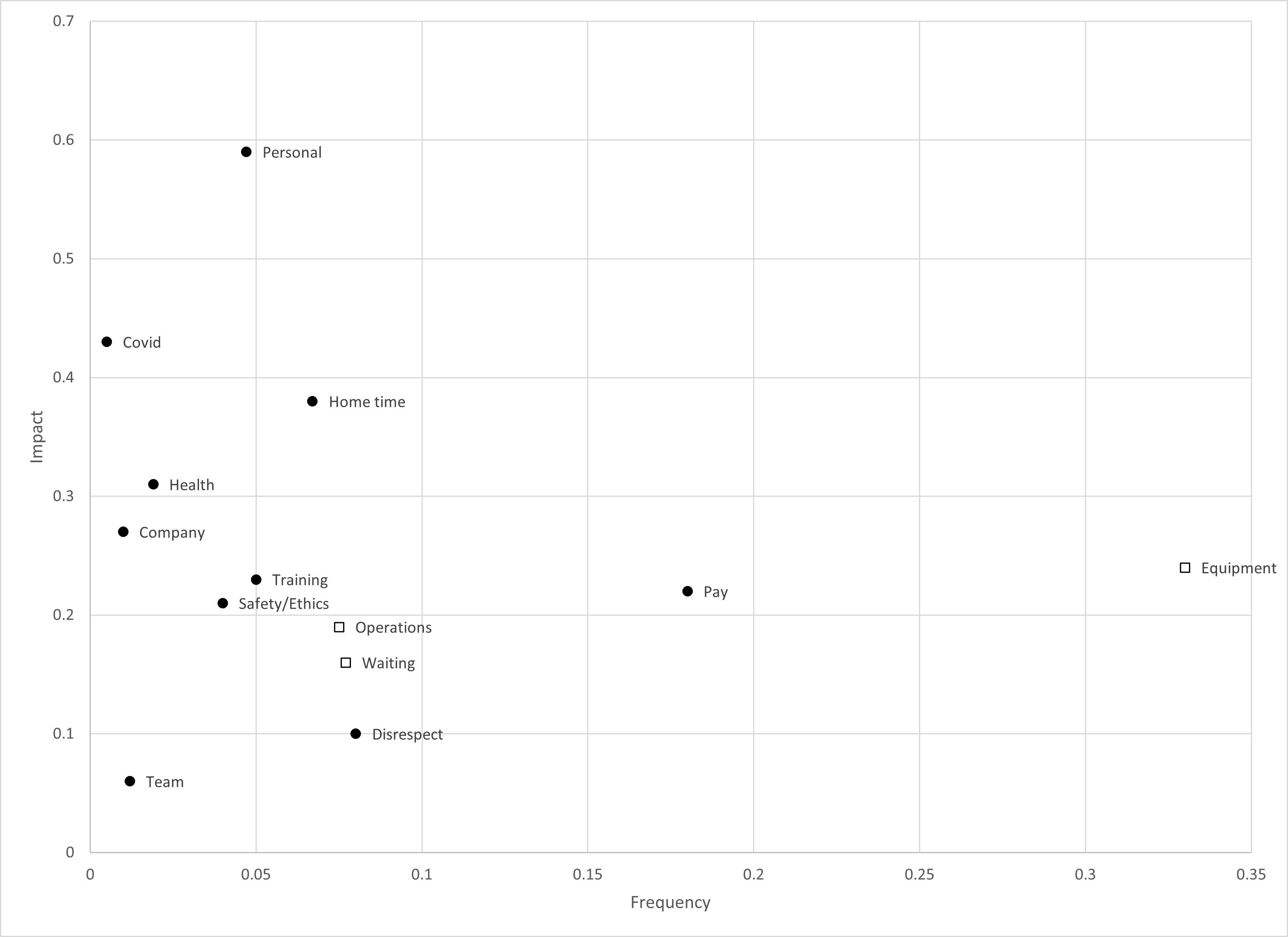}
\end{figure}

To provide more general guidance, in Figure \ref{fg4}, we plot issues by relative frequency and employment impact (based on the coefficients in Table \ref{tb3}), where hollow squares plot the issues that increase the duration of employment and solid black points plot the issues that decrease the duration. We can classify issues into High Frequency \& Medium Impact (Pay, Equipment), Low Frequency \& High Impact (Personal, Covid, Home time, Health, Company) and Low Frequency \& Low Impact (Training, Safety/Ethics, Operations, Waiting, Team, and Disrespect).  Issues in the high-impact category are mostly idiosyncratic to drivers, company issues being a notable exception.  On the other hand, equipment issues stand out as the highest frequency event, and an event that increases employment duration. Timely response to equipment-related concerns may be the best way for companies to reduce turnover.

\bigskip \centerline{\lbrack === Figure \ref{fg4} here ===]}\bigskip 

\subsection{Future Research Directions}
As with all archival research, there is potential for unobserved variables to confound estimates. In this study, the key worry is that unobserved driver type is correlated with communication frequency. While we have undertaken extensive robustness checks to ensure the reliability of our results, the ideal means to fully address the potential for confounding variation would be through a series of field experiments. 


Based on our results, companies in high-turnover industries may consider performing experiments with on-boarding and orientation programs. These programs need not be entirely pleasant experiences, so long as they highlight job-embeddedness characteristics, and perhaps build group unity and identity. Randomly allocating employees to such experiments would allow researchers to isolate any confounding effects while examining the main effects that we find in this paper - shared hardships strengthen the bonds of commitment.


\newpage
\bibliographystyle{agsm}
\bibliography{References_driver_retention}

\newpage
\noindent \large \underline{\textbf{Tables}} \normalsize

\begin{sidewaystable}
\centering
\begin{threeparttable}
\caption{Description of Issue Codes} \label{tb1}
\begin{tabular}{c:l:cc:rr}
\hline
\hline
 &                                                                                                         & \textbf{Turnover} & \textbf{Incentive} &  &     \\
\textbf{Category} &    \textbf{Description}                                                                    & \textbf{Path} & \textbf{Aligned} & \textbf{Freq.} & \textbf{\%}      \\ \hline  
&&&&& \\ [-2mm]
Home time      & Too little home time, home time not as expected, or scheduling change                                                    & 1, 2, 4       & No                & 2,898     & 6.7\%  \\  [2mm]
Waiting        & Poor scheduling/planning, loading delays,   waiting on assigned truck        & 1, 2, 4       & Yes               & 3,347     & 7.7\%   \\ [2mm]
Pay            & Too few or inconsistent miles, rate not competitive, pay inaccurate                                                      & 3             & No                & 7,840     & 18.1\%  \\ [2mm]

Equipment      & Tractor/trailer issues, i.e., breakdowns, maintenance, missing parts                                          & 1, 2, 3, 4    & Yes               & 14,547    & 33.7\%  \\ [2mm]
Safety/Ethics  & Rule violations, unsafe driving/loading conditions, unethical practices                                       & 1             & No                & 1,710     & 4.0\%   \\ [2mm]
Training       & Training pay, accommodations, travel, duration, content                                                                   & 1, 2          & No                & 1,799     & 4.2\%   \\ [2mm]
Team           & Personality conflicts, aversion to team lifestyle                                                                        & 1, 2, 4       & No                & 527       & 1.2\%   \\ [2mm]
Operations     & Routing, loading, fueling, trailer pick-up and drop off                                                                  & 1, 2, 4       & Yes               & 3,245     & 7.5\%   \\ [2mm]
Disrespect     & Driver treatment by dispatchers, motor carrier \& customer personnel                                                    & 1, 2, 4       & No                & 3,672     & 8.5\%   \\ [2mm]
Company        & Company-specific policies, e.g., speed limiters, driver facing cameras,  & 1, 2, 4       & No                & 452       & 1.0\%   \\ [2mm]
        &  sleep apnea machines; company culture conflicts &       &                &       &    \\ [2mm]
Covid          & Issues related to Covid, and Covid-related closures                                                                      & 1             & No                & 224       & 0.5\%   \\ [2mm]
Personal       & Commute to work, retirement, owner- operator preference                                                                   & 1, 2, 4       & No                & 2,028     & 4.7\%   \\ [2mm]
Health         & Non-covid issues related to driver health                                                                                & 1             & No                & 832       & 1.9\%   \\ [2mm]
Others         & Other issues, e.g., no show                                                                                    &               & No                & 77        & 0.2\%   \\ [2mm]
         &                                                                                                                          &               &       & 43,198    & 100.0\% \\ [2mm]
\hline
\hline
\end{tabular}
\end{threeparttable}
\end{sidewaystable}

\vspace{10mm}

\begin{table}[h!]
\centering
\begin{threeparttable}
\caption{Effects of Aligned and Misaligned Issues on Turnover}\label{tb2}
\begin{tabular}{c|ccc}
           & Model 1    & Model 2    & Model 3    \\ \hline \hline
Aligned    & -0.26      & -0.26      & -0.26      \\
(s.e.)       & (0.02)       & (0.02)       & (0.02)       \\
$[$t-stat$]$     & $[$-14.68$]$     & $[$-14.77$]$     & $[$-14.54$]$     \\
\{h.r.\}    & \{-0.23\}      & \{-0.23\}      & \{-0.23\}    \vspace{2mm}  \\
Misaligned & 0.34       & 0.34       & 0.34       \\
           & (0.01)       & (0.01)       & (0.01)       \\
           & $[$24.90$]$      & $[$25.30$]$      & $[$25.01$]$      \\
           & \{0.40\}       & \{0.41\}       & \{0.40\}      \vspace{2mm} \\ \hdashline
Nonresponsive & 0.67       & 0.63       & 0.64       \\
           & (0.06)       & (0.06)       & (0.06)       \\
           & $[$11.95$]$      & $[$11.54$]$      & $[$11.67$]$    \vspace{2mm}   \\ \hdashline
Log-L      & -211,023.10 & -211,184.75 & -181,980.67 \\ \hline
\end{tabular}
\begin{tablenotes}
\item \footnotesize Note: The sample includes 466,236 communication records from 45,873 spells. All specifications include fixed effects for year, month, retention program, call-interval and agent, and a control quantifying the responsiveness of each driver to the human resource company's calls (Nonresponsive = the proportion of calls not answered). Model 1 includes carrier fixed effects, Model 2 includes carrier-level variables, and Model 3 allows for carrier-specific baseline hazard functions. The estimate for each issue type is the average of the estimates for its original issue dummy and four lags. ``h.r." are hazard ratios computed from $\exp(\hat{\beta}) - 1$. Standard errors are clustered at the spell level.\\
\end{tablenotes}
\end{threeparttable}
\end{table}

\clearpage

\begin{table}[h!]
\centering
\begin{threeparttable}
\caption{Effects of Fourteen Major Issue Types on Turnover} \label{tb3}
\centering
\begin{tabular}{c:ccc|c:ccc}
       Category       & Model 1 & Model 2 & Model 3 & Category & Model 1 & Model 2 & Model 3\\ \hline \hline
Home-time     & 0.38   & 0.38   & 0.38   & Operations & -0.19  & -0.20  & -0.19  \\
 (s.e.)             & (0.03)   & (0.03)   & (0.03)   &            & (0.05)   & (0.05)   & (0.05)   \\
 $[$t-stat$]$             & $[$10.89$]$  & $[$11.00$]$  & $[$10.95$]$  &            & $[$-4.14$]$  & $[$-4.22$]$  & $[$-4.09$]$ \\
   \{h.r.\}       & \{0.46\}   & \{0.46\}   & \{0.46\}   &            & \{-0.17\}  & \{-0.18\}  & \{-0.17\}   \\ [1mm]
Waiting       & -0.16  & -0.16  & -0.16  & Disrespect & 0.10   & 0.10   & 0.10   \\
& (0.04)   & (0.04)   & (0.04)   &            & (0.03)   & (0.03)   & (0.03)   \\
& $[$-3.79$]$  & $[$-3.86$]$  & $[$-3.69$]$  &            & $[$3.03$]$   & $[$3.00$]$   & $[$3.03$]$   \\ 
& \{-0.15\}  & \{-0.15\}  & \{-0.14\}  &            & \{0.11\}   & \{0.11\}   & \{0.11\}   \\[1mm]
Pay           & 0.22   & 0.22   & 0.22   & Company    & 0.27   & 0.26   & 0.28   \\
& (0.02)   & (0.02)   & (0.02)   &            & (0.09)   & (0.09)   & (0.09)   \\
& $[$9.41$]$   & $[$9.44$]$   & $[$9.60$]$   &            & $[$3.07$]$   & $[$2.94$]$   & $[$3.27$]$   \\ 
& \{0.24\}   & \{0.24\}   & \{0.25\}   &            & \{0.31\}   & \{0.30\}   & \{0.33\}   \\ [1mm]
Equipment     & -0.24  & -0.24  & -0.24  & Covid      & 0.43   & 0.44   & 0.41   \\
& (0.02)   & (0.02)   & (0.02)   &            & (0.13)   & (0.13)   & (0.14)   \\
& $[$-11.46$]$ & $[$-11.48$]$ & $[$-11.35$]$ &            & $[$3.18$]$   & $[$3.26$]$   & $[$3.06$]$   \\ 
& \{-0.22\}  & \{-0.22\}  & \{-0.21\}  &            & \{0.53\}   & \{0.55\}   & \{0.51\}   \\ [1mm]
Safety/Ethics & 0.21   & 0.23   & 0.20   & Personal   & 0.59   & 0.59   & 0.59   \\
& (0.05)   & (0.05)   & (0.05)   &            & (0.04)   & (0.04)   & (0.04)   \\
& $[$4.47$]$   & $[$4.96$]$   & $[$4.25$]$   &            & $[$14.78$]$  & $[$14.74$]$  & $[$14.72$]$  \\ 
& \{0.23\}   & \{0.26\}   & \{0.22\}   &            & \{0.81\}   & \{0.81\}   & \{0.81\}   \\ [1mm]
Training      & 0.23   & 0.23   & 0.22   & Health     & 0.31   & 0.32   & 0.30   \\
& (0.04)   & (0.04)   & (0.04)   &            & (0.06)   & (0.06)   & (0.06)   \\
& $[$5.21$]$   & $[$5.13$]$   & $[$4.88$]$   &            & $[$5.15$]$   & $[$5.40$]$   & $[$4.87$]$   \\
& \{0.26\}   & \{0.26\}   & \{0.24\}   &            & \{0.36\}   & \{0.38\}   & \{0.35\}   \\  [1mm]
Team          & 0.06   & 0.07   & 0.05   & Others     & 1.10   & 1.06   & 1.15   \\
& (0.09)   & (0.09)   & (0.09)   &            & (0.21)   & (0.19)   & (0.17)   \\
& $[$0.64$]$   & $[$0.80$]$   & $[$0.56$]$   &            & $[$5.26$]$   & $[$5.47$]$   & $[$6.70$]$  \\ 
& \{0.06\}   & \{0.07\}   & \{0.05\}   &            & \{1.99\}   & \{1.89\}   & \{2.17\}  \\ \hline
              \end{tabular}
\begin{tablenotes}
\item \footnotesize Note: The sample includes 466,236 communication records from 45,873 spells. All specifications include fixed effects for year, month, retention program, call-interval and agent, and a control quantifying the responsiveness of each driver to the human resource company's calls (the proportion of calls not answered). Model 1 includes carrier fixed effects, Model 2 includes carrier-level variables, and Model 3 allows for carrier-specific baseline hazard functions. The estimate for each issue category is the average of the estimates for its original issue dummy and four lags. ``h.r." are hazard ratios computed from $\exp(\hat{\beta}) - 1$.
The log-pseudo-likelihood values of Model 1, 2, and 3 are -210,952.6, -211,002.9, and -184,545.8, respectively. Standard errors are clustered at the spell level.
\end{tablenotes}
\end{threeparttable}
\end{table}

\begin{table}[h!]
\centering
\begin{threeparttable}
\caption{Robustness Check (1): Issue Frequencies}\label{tb4}
\begin{tabular}{c|ccc}
           & Model 1    & Model 2    & Model 3    \\ \hline \hline
Aligned                            & -0.24      & -0.24      & -0.24      \\
(s.e.)                               & (0.02)       & (0.02)       & (0.02)       \\
$[$t-stat$]$                             & $[$-12.35$]$     & $[$-12.45$]$     & $[$-12.40$]$     \\
\{h.r.\}                            & \{-0.21\}      & \{-0.21\}      & \{-0.21\}     \vspace{2mm} \\
Misaligned                         & 0.22       & 0.22       & 0.22       \\
                                   & (0.02)       & (0.02)       & (0.02)       \\
                                   & $[$13.84$]$      & $[$13.80$]$      & $[$13.76$]$      \\
                                   & \{0.25\}       & \{0.25\}       & \{0.24\}      \vspace{2mm} \\ \hdashline
Aligned\_f                         & -0.35      & -0.35      & -0.35      \\
                                   & (0.05)       & (0.05)       & (0.05)       \\
                                   & $[$-7.32$]$      & $[$-7.38$]$      & $[$-7.42$]$      \\ [1mm]
(Aligned\_f)$^2$    & 53.08      & 54.95      & 56.83      \\
                                   & (5.31)       & (5.46)       & (5.53)       \\
                                   & $[$10.00$]$      & $[$10.07$]$      & $[$10.28$]$      \\ [1mm]
Misaligned\_f                      & 1.56       & 1.70       & 1.60       \\
                                   & (0.22)       & (0.22)       & (0.22)       \\
                                   & $[$7.21$]$       & $[$7.61$]$       & $[$7.32$]$       \\ [1mm]
(Misaligned\_f)$^2$ & 0.29       & 0.27       & 0.30       \\
                                   & (0.08)       & (0.07)       & (0.08)       \\
                                   & $[$3.88$]$       & $[$3.84$]$       & $[$3.96$]$      \vspace{2mm} \\ \hline
Log-L                              & -210,657.55 & -210,810.14 & -181,620.41 \\ \hline
\end{tabular}
\begin{tablenotes}
\item \footnotesize Note: Aligned\_f and Misaligned\_f are the number of each driver's Aligned or Misaligned issues reported throughout their sample period divided by the period length. Other than these variables and their squared terms, the sample and model specifications are the same as those in Table \ref{tb2}.\\
\end{tablenotes}
\end{threeparttable}
\end{table}

\begin{table}[h!]
\centering
\begin{threeparttable}
\caption{Distribution of Issue Codes under New Categories} \label{tb5}
\begin{tabular}{ccccccccc}
\hline
\hline
Category  & Total   & B-only & OA-only & MA-only & B \& OA & B \& MA & OA \& MA & All three \\ \hdashline
frequency & 35,634   & 7,177   & 6,828    & 15,708   & 1,977    & 749    & 2,803     & 392\\
\%        & 100.0\% & 20.1\% & 19.2\%  & 44.1\%  & 5.5\%   & 2.1\%   & 7.9\%    & 1.1\%  \\  
\hline
\hline
\end{tabular}
\begin{tablenotes}
\item \footnotesize B: Broken-equipment, OA: Other aligned, MA: Misaligned.\\
\end{tablenotes}
\end{threeparttable}
\end{table}

\clearpage

\begin{table}[h!]
\centering
\begin{threeparttable}
\caption{Robustness Check (2): Broken-Equipment Shocks}\label{tb6}
\begin{tabular}{c|ccc}
                         & Model 1    & Model 2    & Model 3 \\ \hline \hline
Broken-equipment only             & -0.42      & -0.43      & -0.42      \\
(s.e.)                     & (0.04)       & (0.04)       & (0.04)       \\
$[$t-stat$]$                   & $[$-11.35$]$     & $[$-11.54$]$     & $[$-11.37$]$     \\
\{h.r.\}                  & \{-0.34\}      & \{-0.35\}      & \{-0.34\}      \\ [1mm]
Other aligned only     & -0.25      & -0.25      & -0.25      \\
                         & (0.03)       & (0.03)    & (0.03)       \\
                         & $[$-7.74$]$      & $[$-7.62$]$      & $[$-7.67$]$      \\
                         & \{-0.22\}      & \{-0.22\}      & \{-0.22\}      \\ [1mm]
Misaligned only         & 0.32       & 0.33       & 0.32       \\
                         & (0.01)       & (0.01)       & (0.01)       \\
                         & $[$21.72$]$      & $[$22.07$]$      & $[$21.79$]$      \\
                         & \{0.38\}       & \{0.39\}       & \{0.38\}       \\ [1mm]
Broken \& Other aligned     & -0.26      & -0.26      & -0.26      \\
                         & (0.06)       & (0.06)       & (0.06)       \\
                         & $[$-4.36$]$      & $[$-4.34$]$      & $[$-4.31$]$      \\
                         & \{-0.23\}      & \{-0.23\}      & \{-0.23\}      \\ [1mm]
Broken \& Misaligned       & 0.20       & 0.19       & 0.20       \\
                         & (0.08)       & (0.08)       & (0.08)       \\
                         & $[$2.57$]$       & $[$2.55$]$       & $[$2.65$]$       \\
                         & \{0.22\}       & \{0.21\}       & \{0.22\}       \\ [1mm]
Other aligned \& Misaligned & 0.15       & 0.15       & 0.16       \\
                         & (0.04)       & (0.04)       & (0.04)       \\
                         & $[$3.99$]$       & $[$4.02$]$       & $[$4.09$]$       \\
                         & \{0.17\}       & \{0.17\}       & \{0.17\}       \\ [1mm]
All three                 & 0.29       & 0.30       & 0.30       \\
                         & (0.10)       & (0.10)       & (0.10)       \\
                         & $[$2.86$]$       & $[$2.97$]$       & $[$2.99$]$       \\
                         & \{0.34\}       & \{0.35\}       & \{0.36\}       \\ [1mm] \hdashline
Log-L                    & -210,981.32 & -211,142.70 & -181,938.63 \\ \hline
\end{tabular}
\begin{tablenotes}
\item \footnotesize Note: The sample includes 466,236 communication records from 45,873 spells. All specifications include fixed effects for year, month, retention program, call-interval and agent, and a control quantifying the responsiveness of each driver to the human resource company's calls. Model 1 includes carrier fixed effects, Model 2 includes carrier-level variables, and Model 3 allows for carrier-specific baseline hazard functions. The estimate for each issue category is the average of the estimates for its original issue dummy and four lags. ``h.r." are hazard ratios computed from $\exp(\beta) - 1$. Standard errors are clustered at the spell level.
\end{tablenotes}
\end{threeparttable}
\end{table}

\clearpage

\begin{table}[h!]
\centering
\begin{threeparttable}
\caption{Robustness Check (3): Broken-Equipment Shocks by Equipment Ownership Status}\label{tb7}
\begin{tabular}{c|cc}
                         & Lease\% \textgreater{} 40 & Lease\% \textless{} 5 \\ \hline \hline
Broken-equipment only             & -0.16                    & -0.37               \\
(s.e.)                     & (0.09)                     & (0.10)                \\
$[$t-stat$]$                   & $[$-1.85$]$                    & $[$-3.67$]$               \\
\{h.r.\}                   & \{-0.15\}                    & \{-0.31\}               \\ [1mm]
Other aligned only     & 0.03                     & -0.17               \\
                         & (0.07)                     & (0.08)                \\
                         & $[$0.47$]$                     & $[$-2.06$]$               \\
                         & \{0.03\}                     & \{-0.16\}               \\ [1mm]
Misaligned only         & 0.41                     & 0.53                \\
                         & (0.04)                     & (0.05)                \\
                         & $[$10.51$]$                    & $[$11.32$]$               \\
                         & \{0.50\}                     & \{0.70\}                \\ [1mm]
Broken \& Other aligned     & -0.14                    & -0.16               \\
                         & (0.14)                     & (0.18)                \\
                         & $[$-1.05$]$                    & $[$-0.91$]$               \\
                         & \{-0.13\}                    & \{-0.15\}               \\ [1mm]
Broken \& Misaligned       & 0.11                     & 0.37                \\
                         & (0.23)                     & (0.18)                \\
                         & $[$0.45$]$                     & $[$2.06$]$                \\
                         & \{0.11\}                     & \{0.45\}                \\ [1mm]
Other aligned \& Misaligned & 0.24                     & 0.33                \\
                         & (0.09)                     & (0.11)                \\
                         & $[$2.56$]$                     & $[$2.95$]$                \\
                         & \{0.27\}                     & \{0.39\}                \\ [1mm]
All three                 & 0.39                     & 0.90                \\
                         & (0.26)                     & (0.22)                \\
                         & $[$1.51$]$                     & $[$4.10$]$                \\
                         & \{0.47\}                     & \{1.45\}                \\ [1mm] \hdashline
\#Observations                    & 51,909                   & 53,123              \\
(\#Spells, \#Carriers)               & (4,819, 6)                    & (4,692, 10)               \\                         \hline
\end{tabular}
\begin{tablenotes}
\item \footnotesize Note: 
Lease rate is calculated by the proportion of leased tractors for each carrier. The two sub-samples, Lease\%$>$40 and Lease\%$<$5, include spells from the carriers with the average lease rate higher than 40\% for years 2019-2021 and from those with the lease rate lower than 5\%. Each sub-sample accounts for about 10\% of the whole sample. The results in this table are computed using the Model 3 in Table 
\ref{tb6}.
\end{tablenotes}
\end{threeparttable}
\end{table}

\begin{table}[h!]
\centering
\begin{threeparttable}
\caption{Distribution of Issue Codes by Equipment Ownership Status} \label{tb8}
\begin{tabular}{lccccccccc}
\hline
\hline
Lease\% $>$ 40 & Total   & B-only & OA-only & MA-only & B \& OA & B \& MA & OA \& MA & All three \\ \hdashline
 & 5,388   & 867   & 1,250    & 2,189   & 374    & 108    & 527     & 73\\
        & 100.0\% & 16.1\% & 23.2\%  & 40.6\%  & 6.9\%   & 2.0\%   & 9.8\%    & 1.4\%  \\
Lease\% $<$ 5 & Total   & B-only & OA-only & MA-only & B \& OA & B \& MA & OA \& MA & All three \\ \hdashline
 & 4,247   & 1,091   & 832    & 1,601   & 247    & 115    & 318     & 43\\
 & 100.0\% & 25.7\% & 19.6\%  & 37.7\%  & 5.8\%   & 2.7\%   & 7.5\%    & 1.0\%  \\
\hline
\hline
\end{tabular}
\begin{tablenotes}
\item \footnotesize B: Broken-equipment, OA: Other aligned, MA: Misaligned.\\
\end{tablenotes}
\end{threeparttable}
\end{table}

\newpage
\noindent \large \underline{\textbf{Figures}} \normalsize

\clearpage

\appendix
\setcounter{table}{0}
\renewcommand{\thetable}{A\arabic{table}}
\textbf{\section*{Appendices}}
\addcontentsline{toc}{section}{Appendices}
\renewcommand{\thesubsection}{\Alph{subsection}}

\section{The Full Sentiment Regression Result}\label{ap2}
The outcome variable is the VADER (Valence Aware Dictionary and sEntiment Reasoner) scores for the communication records. The regressions include the same sets of covariates and fixed effects as in the models in Table \ref{tb6}. Standard errors are clustered at the spell level.

\begin{table}[h!]
\centering
\begin{threeparttable}\caption{Change of Conversation Sentiment after Driver Issues} \label{tb-a2} 
\begin{tabular}{c|ccc} 
Category                & coeff. & s.e.  & t-stat \\ \hline \hline
Broken-equipment only       &        &       &         \\
t=0                         & -0.063 & 0.006 & -10.550 \\
t=1                         & -0.002 & 0.005 & -0.500  \\
t=2                         & 0.010  & 0.005 & 2.170   \\
t=3                         & 0.008  & 0.005 & 1.580   \\
t=4                         & 0.012  & 0.005 & 2.400   \\ \hdashline
Other aligned only          &        &       &         \\
t=0                         & -0.069 & 0.006 & -10.950 \\
t=1                         & -0.013 & 0.005 & -2.670  \\
t=2                         & -0.003 & 0.005 & -0.630  \\
t=3                         & -0.016 & 0.005 & -3.100  \\
t=4                         & -0.003 & 0.005 & -0.640  \\ \hdashline
Misaligned only             &        &       &         \\
t=0                         & -0.060 & 0.004 & -13.840 \\
t=1                         & -0.026 & 0.003 & -7.840  \\
t=2                         & -0.023 & 0.003 & -6.560  \\
t=3                         & -0.016 & 0.004 & -4.400  \\
t=4                         & -0.012 & 0.004 & -3.050  \\ \hdashline
Broken \& Other aligned     &        &       &         \\
t=0                         & -0.244 & 0.013 & -19.150 \\
t=1                         & -0.032 & 0.009 & -3.420  \\
t=2                         & -0.018 & 0.009 & -1.940  \\
t=3                         & -0.028 & 0.010 & -2.870  \\
t=4                         & -0.012 & 0.010 & -1.250  \\ \hdashline
Broken \& Misaligned        &        &       &         \\
t=0                         & -0.249 & 0.022 & -11.560 \\
t=1                         & -0.036 & 0.015 & -2.310  \\
t=2                         & -0.034 & 0.016 & -2.140  \\
t=3                         & -0.014 & 0.017 & -0.850  \\
t=4                         & -0.021 & 0.018 & -1.180  \\ \hdashline
Other aligned \& Misaligned &        &       &         \\
t=0                         & -0.223 & 0.011 & -19.700 \\
t=1                         & -0.051 & 0.008 & -6.020  \\
t=2                         & -0.033 & 0.008 & -3.930  \\
t=3                         & -0.031 & 0.009 & -3.600  \\
t=4                         & -0.025 & 0.010 & -2.640  \\ \hdashline
All three                   &        &       &         \\
t=0                         & -0.391 & 0.031 & -12.420 \\
t=1                         & -0.084 & 0.022 & -3.750  \\
t=2                         & -0.058 & 0.024 & -2.410  \\
t=3                         & -0.055 & 0.024 & -2.330  \\
t=4                         & -0.023 & 0.025 & -0.940  \\ \hdashline
Nonresponsive               & -0.158 & 0.004 & -38.430 \\ \hline
\end{tabular}
\end{threeparttable}
\end{table}

\end{document}